\documentclass[a4paper]{jpconf}
\usepackage{graphicx}
\usepackage{bm}
\newcommand{\bra}[1]{\left\langle #1\right|}
\newcommand{\ket}[1]{\left| #1\right\rangle}
\usepackage{amsfonts}
\begin{document}
\title{A further update on possible crises in nuclear-matter theory}

\author{W. H. Dickhoff}

\address{Department of Physics, Washington University, St. Louis, Missouri 63130, USA}

\ead{wimd@wuphys.wustl.edu}

\begin{abstract}
The ancient problem of the saturation of symmetric nuclear matter is reviewed with an update on the status of the crises that were identified at an early stage by John Clark. 
We discuss how the initial problem with variational calculations providing more binding than the two hole-line contribution for the same interaction was overcome by calculations including three hole-line contributions without however reproducing the empirical nuclear saturation properties.
It is argued that this remaining problem is still open because many solutions have been proposed or ad hoc adjustments implemented without generating universal agreement on the proper interpretation of the physics.
The problem of nuclear saturation therefore persists leading to the necessity of an analysis of the way the nuclear saturation properties are obtained from experimental data.
We clarify the role of short-range correlations and review results for nuclear saturation when such ingredients are completely taken into account using the Green's function method.
The role of long-range correlations is then analyzed with special emphasis on the importance of attractive pion-dominated excitation modes which inevitably lead to higher saturation densities than observed.
Because such modes have no counterpart in finite nuclear systems, it is therefore argued that they should not be considered when calculating nuclear matter properties.
The remaining open question is then whether long-range correlations in finite nuclei which in turn have no counterpart in infinite matter, represent the remaining missing ingredient in this analysis.
We also briefly comment on the role of three-body interactions in the context of the dispersive optical model description of experimental data. 
It is further noted that interactions based on chiral perturbation theory at present do not generate a sufficient number of high-momentum nucleons leading to radii that are too small and substantial overbinding in finite nuclei.
\end{abstract}

\section{Introduction}
On the occasion of the 80th birthday celebration of John Clark it is pertinent to recall his important contributions to the problem of the saturation properties of nuclear matter.
For a more detailed summary of John Clark's contributions to many-body theory the laudatio for the award of the second Feenberg Medal is recommended reading~\cite{Bishop88}. 
Particularly relevant for the present discussion is his summary of the first many-body conference~\cite{Clark79}.
This meeting was dominated by presentations addressing aspects of the nuclear saturation issue.
Variational methods had been developed~\cite{bcc69} that demonstrated a substantial discrepancy with lowest-order Brueckner theory~\cite{Day67} employing the same interaction~\cite{Clark79r,Wiringa79}.
The disagreement between different many-body techniques represented the first crisis identified by John Clark in Ref.~\cite{Clark79}.
This crisis made it clear that the next term in the so-called hole-line expansion~\cite{Day67,Bethe67,Bethe71} involving three independent hole lines could clarify and resolve this situation.
The first implementation of this difficult calculation was accomplished by Day in Refs.~\cite{Day78,Day81,Day81a}.
Confirmation of the compatibility of different methods in establishing the equation of state was reported in Ref.~\cite{Day85}.
Further clarification of the convergence of the hole-line expansion was developed by the Catania group~\cite{Song98} as will be discussed in more detail in Sec.~\ref{sec:hole-line}.

The resolution of this first crisis implied that for nucleon-nucleon (NN) interactions with a substantial repulsive core it is possible to determine the equation of state under the assumption that nucleons are treated nonrelativistically while interacting only by means of two-body interactions.
Added to these assumptions should be the notion that there is a limitation on the density range where these statements are valid.
This limitation can be linked to the convergence criterion applied to the hole-line expansion and essentially translates into the notion that nuclear matter should still be a relatively low-density system~\cite{Day67}.
Unfortunately, the resolution of the first crisis was followed by a second crisis as identified by John Clark~\cite{Clark79} namely that nuclear-matter saturation properties were not explained in detail by the assumptions of nonrelativistic nucleons interacting by means of two-body interactions only.
All results remained substantially removed from the empirical density of 0.16 nucleons per fm$^{3}$ and binding energy of 16 MeV per particle mostly in such a way that the binding energy is described correctly but at too high a saturation density.
The latter systematics represent an improvement compared to the much more distinct Coester band obtained in lowest-order Brueckner theory for different interactions, where the location of the saturation point is related to the amount of $D$-state probability the interaction generates, large values leading to lower saturation densities and \textit{vice versa}~\cite{Coester70}. 

The resulting second crisis in nuclear-matter theory has been studied by many people and a multitude of remedies have been proposed.
Since no uniform agreement on its resolution has emerged, this crisis endures to this day and several of its aspects will be discussed in the following.
Section~\ref{sec:hole-line} is devoted to a discussion of some salient features of the three hole-line results of Ref.~\cite{Song98} together with some standard remedies for solving the discrepancy with the empirical saturation properties.
Section~\ref{sec:finiten} is devoted to a brief summary of what has been learned from Green's function Monte Carlo calculations in light nuclei~\cite{Pieper01} and recent developments that constrain the nucleon self-energy in ${}^{40}$Ca~\cite{Mahzoon14,Dussan14} including its high-momentum content.
Insights from experiments that determine quantities relevant for saturation are discussed on the basis of the lessons provided by these considerations of finite nuclei.
In particular, the value of the interior density appear to be dominated by the physics induced by short-range correlations (SRC).
Other ingredients like the deviations of the mean field found in $(e,e'p)$ reactions are put together to provide an alternative perspective of nuclear saturation in Sec.~\ref{sec:nucsatx}.
The impact of recent applications of NN interactions based on chiral perturbation theory is also briefly discussed.
Some final words regarding John Clark's impact are presented in Sec.~\ref{sec:conc}.

\section{Ingredients of the hole-line expansion}
\label{sec:hole-line}
\begin{figure}[tb]
\begin{center}
\includegraphics[angle=0,width=0.9\textwidth]{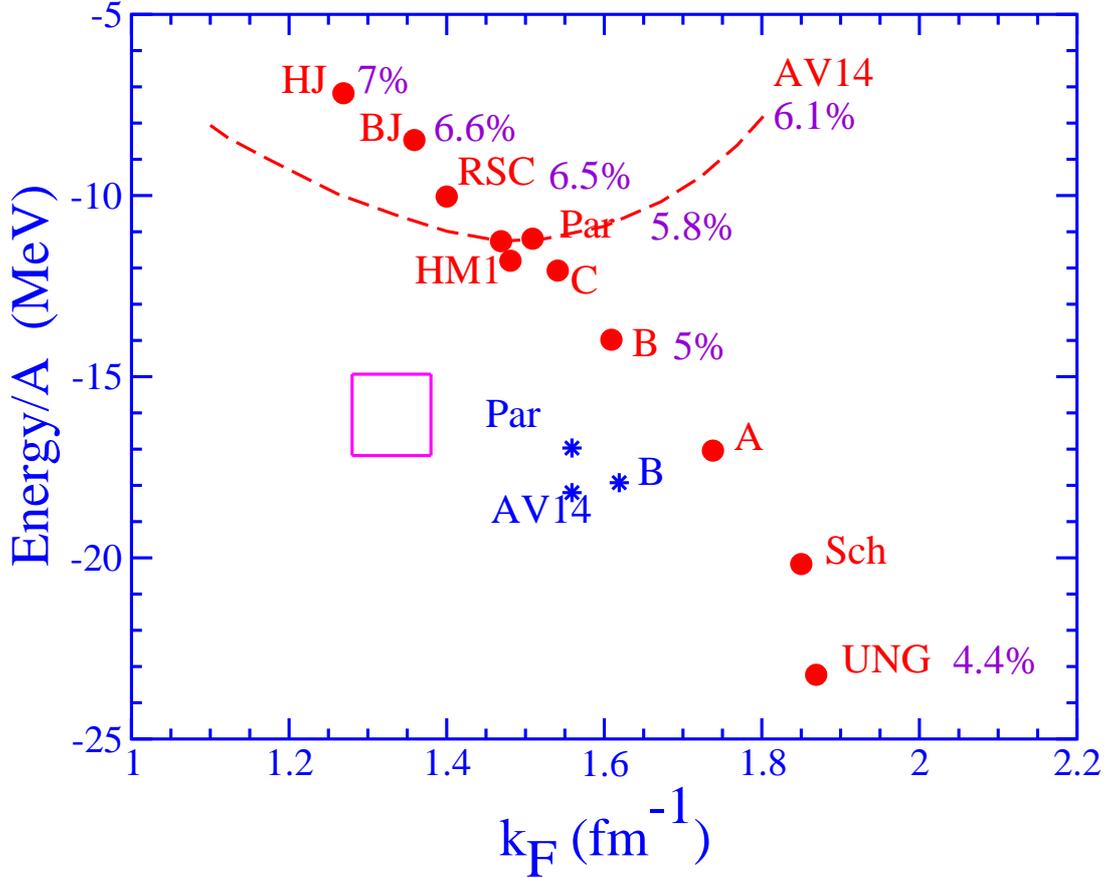}
\caption{Energy per particle as a function of the Fermi wave vector
for different realistic NN interactions.
The circles indicate the minima of the saturation curve for the BHF
approximation when the standard choice of $U$ is made.
The symbols identify the various interactions and in some cases the
$D$-state admixture in the deuteron wave function is also given.
The star symbols are associated with the minima of the saturation curve
when three hole-line contributions are included.
The box identifies the empirical region suggested by experimental data.
For the Argonne $v14$ (AV14) interaction the saturation curve
in the BHF approximation is represented by the dashed line.
The figure has been adapted from Ref.~\protect\cite{Baldo99}.
\label{fig:coesterb}}
\end{center}
\end{figure}

The three hole-line calculation reported in Ref.~\cite{Song98} allowed for the first time an assessment of the choice of the auxiliary potential that can be introduced in the perturbation expansion of the ground-state energy.
Many applications of lowest-order Brueckner theory had adopted the so-called standard choice in which the single-particle spectrum below the Fermi energy is calculated self-consistently from the Brueckner-Hartree-Fock (BHF) self-energy.
The BHF self-energy can be calculated in analogy to the Hartree-Fock contribution but the bare interaction is replaced by Brueckner's $G$-matrix which sums all ladder diagrams while allowing only intermediate particle states corresponding to momenta above the Fermi momentum.
In the standard choice the intermediate particle states in the $G$-matrix propagate with energies that correspond to  kinetic ones reducing the computational effort.
In the so-called continuous choice~\cite{Jeukenne76} the BHF prescription is extended to include momenta above the Fermi momentum as well and requires the recalculation of the $G$-matrix in each iteration step in which the single-particle spectrum is updated.
In addition, only the real part of the BHF self-energy is included in this procedure (for momenta below the Fermi momentum there is no imaginary part).
The three hole-line calculations of Ref.~\cite{Song98} included both choices for the auxiliary potential.
As was well known at the time, the continuous choice provides consistently more binding than the standard choice at the level of two-hole lines.
When three-hole line terms are added in each case the total result was almost indistinguishable at least for densities sufficiently beyond the expected saturation point demonstrating independence of the choice of the auxiliary potential and excellent convergence properties from two to three hole lines as suggested by Bethe.
Furthermore, the continuous choice already approximated this final result reasonably at the two hole-line level allowing for a straightforward calculation of the equation of state at reasonable computational expense.
Results summarizing the situation are illustrated in Fig.~\ref{fig:coesterb} which is adapted from Ref.~\cite{Baldo99}.
The open circles in Fig.~\ref{fig:coesterb} indicate the minimum of the energy per
particle for different realistic NN interactions identified by the
appropriate abbreviations. For the AV14 interaction~\cite{Wiringa84}
the dashed curve gives the relevant part of the complete curve.
All open circles correspond to BHF calculations of the energy per particle confirming the identification of the so-called Coester band~\cite{Coester70} with the location of the minimum governed by the $D$-state probability 
of the deuteron that is generated by the interaction.
Including all three hole-line terms, moves the saturation points
away from the Coester band towards the empirical region,
as illustrated in Fig.~\ref{fig:coesterb} by the stars for three
different realistic interactions~\cite{Baldo99}.
The remaining discrepancy with empirical data is still substantial, however, especially when the results are plotted with respect to the density.

So the nuclear saturation problem remains!
Several remedies have been proposed over the years and we will now consider
some of their features.
The first is closely associated
with the presence of excited states of the nucleon, in particular the 
$\Delta$-isobar.
Its importance in pion-nucleon scattering suggests that it may be necessary to include it on the same
footing as the nucleon.
The disadvantage of this strategy is that a lot of information is required
about the interaction between nucleons and $\Delta$-isobars for which
few experimental constraints are available.
An alternative strategy is to represent the influence of $\Delta$-isobars
and other nucleonic excitations by including three- and perhaps higher-body
interactions.
The occurrence of such interactions is inevitable if one restricts the
quantum Hilbert space to nucleons~\cite{fumi57}.
More elaborate versions of this type of three-body force~\cite{coon79}
yield attractive contributions to the energy per particle.
Since light nuclei require additional binding beyond the contribution of
two-body interactions, such three-body forces help in getting better
agreement for light nuclei.
In nuclear matter, however, the situation is more complicated since
the Coester band properties suggest that a repulsive mechanism is
needed to generate lower saturation densities that are in accord
with the empirical results.
For this reason an additional phenomenological repulsive three-body
interaction was introduced by~\cite{capawi83}, which
was then adjusted to force the correct
saturation properties of nuclear matter, while also fitting the binding
of light nuclei.
The procedure yields an improved Hamiltonian for nuclei but
gives up on a deeper insight into the saturation mechanism of nuclear
matter, especially since the origin of the effective repulsion is somewhat 
unclear.
While the interactions have become more refined, the situation remains as described when two-body NN interactions are employed with strong repulsive cores.

An alternative solution to the saturation problem has been proposed that
includes aspects of the effects of relativity~\cite{acps}.
A detailed discussion is beyond the scope of the present talk
and only a few comments will be given here for completeness.
By employing a straightforward adaptation
of the BHF approach, the so-called Dirac--BHF (DBHF) method gives reasonable
saturation properties for nuclear matter~\cite{hama,brma}.
The main physical effect appears to be the change of the coupling of
the so-called $\sigma$-meson to the nucleons in the medium.
This scalar, isoscalar meson represents to a large extent the physical 
exchange between nucleons of two interacting pions, coupled to zero angular 
momentum and isospin.
Since the actual form of the scalar coupling of the $\sigma$-meson to the 
nucleons is essential in obtaining the saturation mechanism, it is unclear to
what extent it truly represents the two pion-exchange processes in the medium.
An additional difficulty is the necessity to deal with
the properties of antiparticles and the so-called Dirac sea.
Further study of higher-order (three hole-line) contributions have so far 
not taken place to assess the convergence properties of the scheme.

\section{Lessons from finite nuclei}
\label{sec:finiten}
Some properties of light nuclei can nowadays be calculated in an exact manner 
with different techniques starting from a realistic NN interaction.
An example is the application of several methods to the 
calculation of the ground-state energy of ${}^4\mathrm{He}$, reported 
in~\cite{kama01}.
The low-lying states of nuclei up to $A=10$~\cite{pvw02} 
can be described with the Green's function Monte 
Carlo (GFMC) method~\cite{Pieper01}.
This body of work is able to explain many aspects of the low-energy spectra 
of light nuclei, starting from a realistic interaction.
Many details are further improved by including a three-body interaction 
between the nucleons.
In all cases studied so far, the calculated energy of the ground state is 
always above the experimental number when only two-body interactions are 
included: a clear indication for the need of an overall attractive 
three-body force corresponding to about 1.5 MeV per particle attraction for the Argonne calculations~\cite{Pieper01}.
The two-body interaction used in the GFMC method is known as the AV18~\cite{Wiringa1995}.
It contains a substantial repulsive core capable of generating an appropriate amount of short-range correlations~\cite{Wiringa14} demanded by exclusive electron scattering experiments~\cite{Subedi08}.
With additional adjustments of the three-body interactions added to AV18 an excellent description of $p-$shell nuclei is obtained including the ground state of ${}^{12}$C.
We note that the AV18 is a local interaction which facilitates its implementation in the GFMC framework.

The issue of high-momentum components has recently been addressed in the context of developments of the dispersive optical model (DOM) originally developed by Mahaux and Sartor~\cite{Mahaux91}.
In Ref.~\cite{Mahzoon14} a nonlocal representation of the nucleon self-energy for ${}^{40}$Ca was constrained by all available data related to the nucleon propagator including elastic scattering and structure information related to the ground state.
The latter information includes removal of high-momentum nucleons obtained at Jefferson Lab~\cite{Rohe04,Rohe04A}.
\begin{figure}[tbp]
\begin{center}
\includegraphics*[scale=0.75]{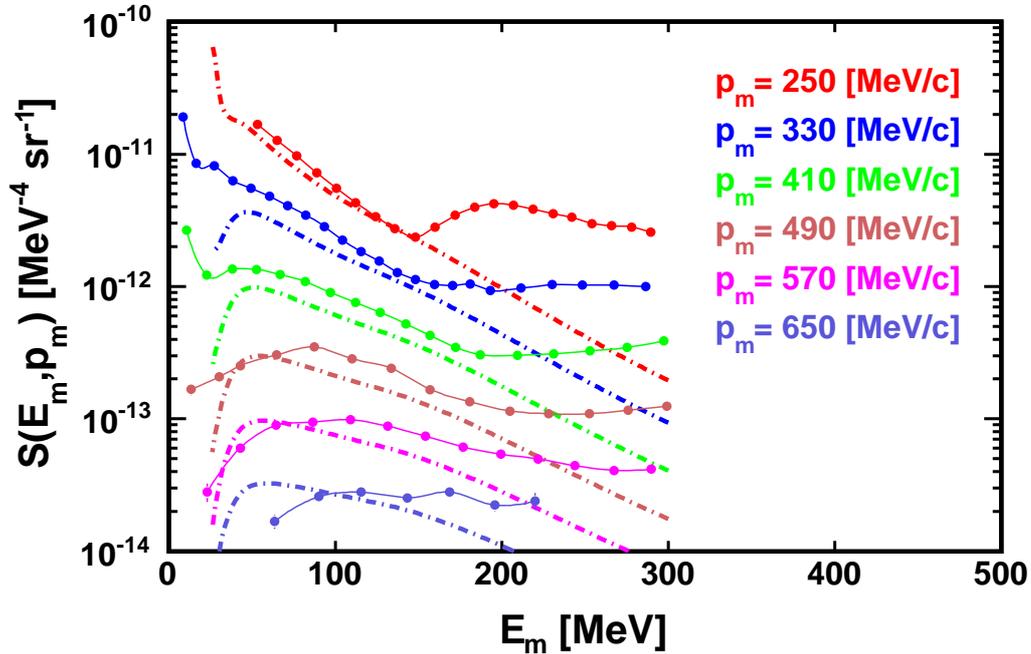}
\caption{Spectral strength as a function missing energy for different missing momenta as indicated in the figure. The data  are the average of the ${}^{27}$Al and ${}^{56}$Fe measurements from Ref.~\protect\cite{Rohe04A}.}
\label{fig:highk}
\end{center}
\end{figure}
In Fig.~\ref{fig:highk} the results for the high-momentum removal spectral strength obtained in Ref.~\cite{Mahzoon14} are compared with the Jefferson Lab data~\cite{Rohe04A}.
The high-energy data correspond to intrinsic nucleon excitations and are not part of the DOM analysis.
To further improve the description, one would have to introduce an energy dependence of the radial form factors for the potentials. Nevertheless an adequate description is generated which corresponds to 10.6\% of the protons having momenta above 1.4 $\textrm{fm}^{-1}$.

The energy sum rule for the ground state~\cite{Dickhoff08} can be expressed as
\begin{eqnarray}
E^N_0  = 
\frac{1}{2\pi} \int_{-\infty}^{\varepsilon_F^-} \!\! dE\
\sum_{\alpha,\beta} \left\{ \bra{\alpha} T \ket{\beta}
+\ E\ \delta_{\alpha,\beta}
\right\} \textrm{ Im } G(\beta,\alpha;E) , 
\label{eq:erule}
\end{eqnarray}
where $G(\beta,\alpha;E)$ represents the single-particle propagator in a general basis.
Equation~(\ref{eq:erule}) is valid when only two-body interactions are present in the Hamiltonian.
In practice it is convenient to perform this calculation in momentum space employing the momentum distribution 
$n_{\ell j}(k)$ and $S_{\ell j}(k;E)$ the spectral function for a given $\ell j$ combination.
The results quoted below are corrected for center-or-mass effects in the form given in Ref.~\cite{Dieperink74}.
A binding energy of 7.91 MeV/$A$ is obtained much closer to the experimental 8.55 MeV/$A$ than found in Ref.~\cite{Dickhoff10a} for a local version of the DOM.
The constrained presence of the high-momentum nucleons is responsible for this change~\cite{Muther95}.
The 7.91 MeV/$A$ binding incorporates the contribution to the ground-state energy from two-body interactions including a kinetic energy of 22.64 MeV/$A$ and was not part of the fit.
This empirical approach therefore leaves about 0.64 MeV/$A$ attraction for higher-body interactions but probably should be accompanied by an error of similar size at this stage. 
This statement requires consideration of the modification of the energy sum rule in the presence of \textit{e.g.} a three-body interaction denoted by $\hat{W}$.
The corresponding result can be expressed in the following form~\cite{Carbone13}
\begin{eqnarray}
E^N_0 
= \frac{1}{2\pi} \int_{-\infty}^{\varepsilon_F^-} \!\! dE\
\sum_{\alpha,\beta} \left\{ \bra{\alpha} T \ket{\beta}
+\ E\ \delta_{\alpha,\beta}
\right\} \textrm{ Im } G(\beta,\alpha;E) -\frac{1}{2} \bra{\Psi^N_0} \hat W \ket{\Psi^N_0} ,
\label{eq:erule3}
\end{eqnarray}
where the last term denotes the explicit contribution of the three-body term.
The form of Eq.~(\ref{eq:erule3}) clarifies that the above analysis for the propagator which in principle includes the effect of three-body interactions, requires a \textit{repulsive} contribution of the three-body expectation value of 1.28 MeV/$A$.
This result should be compared to the approximately 1.5 MeV/$A$ \textit{attraction} needed for light nuclei
in the GFMC calculations of Ref.~\cite{Pieper01} for light nuclei.
The size of the three-body contribution whether attractive or repulsive of about 1.5 MeV/$A$ suggests nevertheless that in comparison with about 30 MeV/$A$ attraction from two-body terms there is a reasonable convergence with respect to the number of interacting bodies for the energy of the ground-state, assuming that the DOM spectral properties determining Eq.~(\ref{eq:erule}) are dominated by two-body contributions.

At this point it is helpful to recall how nuclear saturation properties are determined from experimental data.
With this in mind we can consider whether and how these properties are determined by SRC and perhaps long-range correlations (LRC). 
For the latter consideration it is important to consider LRC in finite nuclei and infinite matter and determine if there is actually any relation between the two.
Arguments are presented in the following that SRC represent the dominant
factor in determining the empirical saturation density of nuclear matter.
We recall that elastic electron scattering from
${}^{208}$Pb~\cite{Frois77} accurately determines
the value of the central charge density.
By multiplying this number by $A/Z$ one obtains the relevant central density.
For this nucleus one then finds 0.16 nucleons/fm${}^3$ or $k_F = 1.33$ fm${}^{-1}$.
The presence of nucleons at the center of a heavy nucleus is confined
to $s$-wave nucleons.
Calculations of momentum distribution in nuclear matter suggest that the depletion of deeply bound states is dominated by SRC~\cite{Vonderfecht91} generating a global depletion of the Fermi sea. 
The deviation from a constant depletion near the Fermi energy is due to the presence of LRC which only play a role in that energy domain.

Experimental confirmation of such features were provided by one of the last experiments at the NIKHEF facility employing the $(e,e'p)$ reaction on ${}^{208}$Pb which covered the whole energy and momentum domain of the mean field for protons~\cite{bat} (see also Ref.~\cite{Dickhoff08}).
Experimental occupation numbers for mean-field proton orbits obtained in Ref.~\cite{bat} demonstrate that also in the finite nucleus a global depletion of the Fermi sea occurs which exhibits only some orbital dependence near the Fermi energy.
It is therefore particularly appropriate to conclude that the depletion of the deeply bound
$0s\scriptstyle \frac{1}{2}$ and $1s\scriptstyle \frac{1}{2}$ protons is determined by SRC roughly corresponding to about 15\% while it also pertains to
a large extent for the $2s\scriptstyle \frac{1}{2}$ orbit at the Fermi energy with only perhaps an additional 10\% due to LRC~\cite{Dickhoff10}.
These considerations demonstrate clearly that one may expect
SRC to have a decisive influence on the actual value
of the nuclear-matter saturation density extracted from finite nuclei.
The above arguments can be applied to neutrons in ${}^{208}$Pb and the $A/Z$ multiplication may be more or less  valid as there are four mostly occupied neutron $s$-orbits. Nevertheless an assumption about the nucleon asymmetry dependence of the saturation density will have to made as well.
More unambiguous is the situation for ${}^{40}$Ca where the interior charge density is expected to be accurately matched by the neutron one~\cite{Mahzoon14} on account of isospin symmetry.
Accordingly, a value of about 0.16 nucleons per fm$^3$ does emerge in this nucleus.

The argument of the dominance of SRC was made for example in Ref.~\cite{Dewulf03} and calculations to completely treat SRC in nuclear matter including full off-shell propagation in solving the ladder equation have been successfully implemented in particular at finite temperature~\cite{Frick05,Rios09} using the self-consistent Green's function (SCGF) method~\cite{Dickhoff04}.
As demonstrated in Ref.~\cite{Dewulf03} for other realistic interactions, the latest nuclear-matter results for the AV18 interaction confirm that only treating SRC but fully self-consistently with results extrapolated to $T=0$ generates saturation at the correct density~\cite{Baldo12}.
The traditional amount of binding that is expected at saturation is obtained by extrapolating the empirical mass formula for $N=Z$ to infinite particle number while discarding the Coulomb contribution.
The resulting value suggests a binding at saturation of about 16 MeV/$A$.
The amount of binding at saturation for AV18 from the SCGF calculation reported in Ref.~\cite{Baldo12} is about 4 MeV short.
This issue will be addressed in the next section.

\section{Alternative perspective of nuclear saturation}
\label{sec:nucsatx}
It remains to
be understood why apparently converged hole-line calculations~\cite{Song98} yield higher saturation densities. The three
hole-line terms obtained in Ref.~\cite{Song98,Baldo12} indicate reasonable
convergence properties compared to the two hole-line
contribution. One may therefore assume that these results
provide an accurate representation of the energy per particle
of nuclear matter as a function of density for the case
of nonrelativistic nucleons and two-body forces. At this
point it is useful to identify a hidden assumption
when the nuclear matter problem is posed.
It asserts that the influence of LRC
in finite nuclei and nuclear matter are commensurate. 
It has been suggested in Ref.~\cite{Dewulf03} that this underlying assumption 
is questionable. 
The argument is based on the special properties of the LRC 
associated with pion-exchange interactions.
These attractive LRC contribute at transferred
wave vectors of about 1 to 2 fm$^{-1}$.
Ring diagram summations of such attractive interactions yield a coherent sum (all
terms are attractive).
For interactions different from the Coulomb one, the integral over this momentum variable $\bm{q}$
generates no contribution to the ground-state energy for small values on account of the $dq~q^2$ term.
The pion-exchange terms do not suffer this fate, since they occur at finite
$q$, and are amplified due to momentum conservation in an infinite system.
It is unclear whether these terms actually generate a similar 
physical consequence in finite
nuclei, where momentum is not a good quantum number.
The nuclear matter response with pion quantum numbers demonstrates that low-lying strength should accumulate.
The lack of such collectivity with pionic quantum numbers in nuclei 
illustrates that the relation between nuclear matter and finite nuclei,
at least for these degrees of freedom, is nonexistent.
Indeed, experimental data exhibit no enhanced response of the pion channel over that
of the rho~\cite{carey}, as demanded by nuclear
matter calculations.
Given this inconsistency, one can conclude that binding-energy contributions
of long-range pion-exchange terms \textbf{do not} play the same role in finite nuclei.
It is therefore necessary to excise them from nuclear-matter 
calculations to establish contact with finite nuclei.
In other words, the original nuclear matter problem has been ill-posed
and only the effects of SRC should be employed
to connect the infinite with the finite system.

Three hole-line contributions do 
include such a third-order ring diagram characteristic of LRC. The corresponding effect of these LRC
on nuclear saturation properties is sizable, as shown by
the results for three- and four-body ring diagrams calculated
in Ref.~\cite{Dickhoff82} (see also Refs.~\cite{Day81,Song98}). The results
of Ref.~\cite{Dickhoff82,Dickhoff84} demonstrate that such ring-diagram terms
are indeed dominated by attractive contributions involving
pion quantum numbers propagating around the rings and
increase substantially in importance with increasing density. 
As has been argued above, these contributions have no counterpart in finite nuclei as they are simply too small. 
Excising the three ring-diagram  contribution from the three hole-line result~\cite{Song98} indeed confirms that a sensible saturation density is obtained.

It is also important to note that 
even with the right nuclear-matter saturation properties nothing is explained if corresponding calculations in finite nuclei generate a nuclear charge density in the interior that is too large.
Furthermore, it is in turn quite clear that LRC which clearly exist in finite nuclei as surface vibrations and giant resonances~\cite{Brand90} 
have no corresponding counterpart in infinite nuclear matter.
Whether such phonon contributions to the binding energy of finite nuclei are completely negligible is also not so obvious but remains an open question.
It is certainly possible that such terms could contribute a few MeV per particle necessary to bridge the missing binding that apparently is generated in nuclear matter when only SRC are (sophisticatedly) included in the SCGF results for AV18 reported in Ref.~\cite{Baldo12}.

We conclude this section with some brief remarks about the use of NN interactions based on chiral perturbation theory.
In practical developments these interactions are cut in momentum space corresponding to about 500 MeV.
If such a cut-off is introduced from the beginning, it is clear that no sizable SRC will be introduced.
Indeed the momentum properties of nucleons in the ground state of nuclei observed experimentally at Jefferson Lab suggest~\cite{Rohe04,Rohe04A,Subedi08,Arrington2011} that an interaction like AV18 is a much more appropriate vehicle to describe these nucleons~\cite{Wiringa14}.
It is therefore not surprising that \textit{ab initio} calculations employing soft chiral interactions generate heavier nuclei that have a too small radius and exhibit substantial overbinding (see \textit{e.g.} Ref.~\cite{Soma14}).
The impossibility to simultaneously to describe nuclear saturation and the triton binding energy with standard versions of chiral interactions is also noteworthy in this context~\cite{Hagen14}.
The need for quite large corrections from three-body forces also appears inconsistent with the GFMC results for light nuclei and the empirical DOM result for ${}^{40}$Ca~\cite{Mahzoon14}.
Finally, it should be observed that lattice calculations that aim to generate the NN interaction directly from QCD~\cite{Ishii07} generate a strongly repulsive core when the pion mass assumes realistic values, thereby providing further indications that future applications to nuclear saturation and finite nuclei should proceed on the basis of NN interactions with strongly repulsive cores.

\section{Dedication}
\label{sec:conc} 
The final words in this contribution should emphasize the importance of the efforts of John Clark in understanding nuclear-matter saturation. They continue to inspire some people like the author to persist in thinking about solutions to this problem in as complete and physically sensible a way as possible.
Some of these ideas have been put forward in the present paper which is dedicated to John on the occasion of his 80th birthday!

\section*{Acknowledgment} This work was supported by the U.S. National Science Foundation under grants PHY-1304242.
\section*{References}
\bibliography{MBT18}

\end{document}